\documentclass[%
 aip,
 jmp,%
 amsmath,amssymb,
 reprint,%
]{revtex4-1}

\usepackage{graphicx}% Include figure files
\usepackage{dcolumn}% Align table columns on decimal point
\usepackage{bm}% bold math
\begin{document}

\preprint{AIP/123-QED}

\title{Cryogenic on-chip multiplexer for the study of quantum transport in 256 split-gate devices\\}% Force line breaks with \\
%\thanks{Footnote to title of article.}

\author{H. Al-Taie}
\email{ha322@cam.ac.uk}
\affiliation{Centre for Advanced Photonics and Electronics, Electrical Engineering Division, Department of Engineering, 9 J. J. Thomson Avenue, University of Cambridge, Cambridge, CB3 0FA, United Kingdom\\}%\\This line break forced with \textbackslash\textbackslash}%
\affiliation{Cavendish Laboratory, Department of Physics, University of Cambridge, J. J. Thomson Avenue, Cambridge, CB3 0HE, United Kingdom\\}%\\This line break forced with \textbackslash\textbackslash}%

\author{L. W. Smith}
\author{B. Xu}
\affiliation{Cavendish Laboratory, Department of Physics, University of Cambridge, J. J. Thomson Avenue, Cambridge, CB3 0HE, United Kingdom\\}%\\This line break forced with \textbackslash\textbackslash}%

\author{P. See}
\affiliation{National Physical Laboratory, Hampton Road, Teddington, Middlesex, TW11 0LW, United Kingdom\\}%\\This line break forced with \textbackslash\textbackslash}%

\author{J. P. Griffiths}
\author{H. E. Beere}
\author{G. A. C. Jones}
\author{D. A. Ritchie}
\affiliation{Cavendish Laboratory, Department of Physics, University of Cambridge, J. J. Thomson Avenue, Cambridge, CB3 0HE, United Kingdom\\}%\\This line break forced with \textbackslash\textbackslash}%

\author{M. J. Kelly}
\affiliation{Centre for Advanced Photonics and Electronics, Electrical Engineering Division, Department of Engineering, 9 J. J. Thomson Avenue, University of Cambridge, Cambridge, CB3 0FA, United Kingdom\\}%\\This line break forced with \textbackslash\textbackslash}%
\affiliation{Cavendish Laboratory, Department of Physics, University of Cambridge, J. J. Thomson Avenue, Cambridge, CB3 0HE, United Kingdom\\}%\\This line break forced with \textbackslash\textbackslash}%

\author{C. G. Smith}
\affiliation{Cavendish Laboratory, Department of Physics, University of Cambridge, J. J. Thomson Avenue, Cambridge, CB3 0HE, United Kingdom\\}%\\This line break forced with \textbackslash\textbackslash}%

\date{\today}% It is always \today, today,
             %  but any date may be explicitly specified            
\begin{abstract}
We present a multiplexing scheme for the measurement of large numbers of mesoscopic devices in cryogenic systems. 
The multiplexer is used to contact an array of 256 split gates on a GaAs/AlGaAs heterostructure, in which each split gate can be measured individually. 
The low-temperature conductance of split-gate devices is governed by quantum mechanics, leading to the appearance of conductance plateaux at intervals of $2e^2/h$.
A fabrication-limited yield of $94 \%$ is achieved for the array, and a `quantum yield' is also defined, to account for disorder affecting the quantum behaviour of the devices. 
The quantum yield rose from $55 \%$ to $86 \%$ after illuminating the sample, explained by the corresponding increase in carrier density and mobility of the two-dimensional electron gas. 
The multiplexer is a scalable architecture, and can be extended to other forms of mesoscopic devices. It overcomes previous limits on the number of devices that can be fabricated on a single chip due to the number of electrical contacts available, without the need to alter existing experimental set ups.

\end{abstract}

%\pacs{Valid PACS appear here}% PACS, the Physics and Astronomy
                             % Classification Scheme.
%\keywords{Multiplexer, Split gate, 1D quantum wires, Mesoscopic devices}%Use showkeys class option if keyword
                              %display desired
\maketitle

There has been much interest in using gate-defined mesoscopic devices for computational applications, from spintronics to quantum information processing. For example, it has recently been shown that the current through a 1D conductor defined using a split gate can be spin polarized by purely electrical means~\cite{Debray2009, Chen2012}. This has great potential in spintronics~\cite{Awschalom2007} where electrical control of the electron spin is highly advantageous~\cite{Awschalom2009}. In addition, much research is focussed on using quantum dot systems as spin and charge qubits~\cite{Loss1998, Hayashi2003, Hanson2007, Kloeffel2013}.

In order for such mesoscopic devices to form the building blocks of integrated quantum circuits, the yield and manufacturability must be considered. The definition of yield for mesoscopic devices will include the reproducibility of quantum phenomena from device to device, as well as the reliability of fabrication processes.
It must also be shown that the devices can be integrated into a scalable architecture. Significant progress has been made in fabricating large-scale arrays of nanowires~\cite{Yao2013} and carbon nanotubes~\cite{Park2012}, but no equivalent attempts have been made for gate-defined structures on GaAs/AlGaAs heterostructures.

We have fabricated a large array of gate-defined mesoscopic devices on the surface of a GaAs/AlGaAs heterostructure. Each device in the array can be measured individually using a multiplexing scheme. To demonstrate the functionality of the multiplexer, we form an array of 256 one-dimensional (1D) quantum wires, defined using split gates \cite{Thornton1986}. The split gate was chosen because it is one of the simplest mesoscopic devices that exhibits quantum phenomena; the quantisation of conductance in units of $2e^2/h$ as a function of split-gate voltage~\cite{vanWees1988, Wharam1988}.
Measuring a large number of devices during a single cooldown in a cryostat enables a systematic study of the yield and the reproducibility of electrical characteristics of the devices. It also provides a data set that is sufficiently large for statistical analysis of quantum phenomenon. The array structure is highly scalable, such that many more devices can be incorporated with few extra electrical contacts.
We have focussed on 1D devices, however, the multiplexer methodology can be extended to arrays of other types of quantum devices, for example quantum dots.

The devices were fabricated on a modulation-doped GaAs/AlGaAs High Electron Mobility Transistor (HEMT) structure, in which the two-dimensional electron gas (2DEG) is formed $90$ nm below the surface of the wafer. Data are presented before and after illumination with a red LED. Before illumination, the carrier density ($n$) and mobility ($\mu$) were measured to be $1.7\times10^{11}$ cm$^{-2}$ and $0.94\times10^6$ cm$^2$V$^{-1}$s$^{-1}$, respectively. After illumination, $n$ and $\mu$ increased to $2.9\times10^{11}$ cm$^{-2}$ and $2.2\times10^6$ cm$^2$V$^{-1}$s$^{-1}$, respectively. Schottky gates on the surface of the sample were patterned using optical lithography, with the exception of the split gates, which were patterned using electron-beam lithography. Each split gate was $400$ nm long, and $400$ nm wide. Two-terminal lock-in measurements were performed at 1.4 K using an ac excitation voltage of 100 $\mu$V.

\begin{figure}
\includegraphics[width=8.5cm,height=10cm,keepaspectratio]{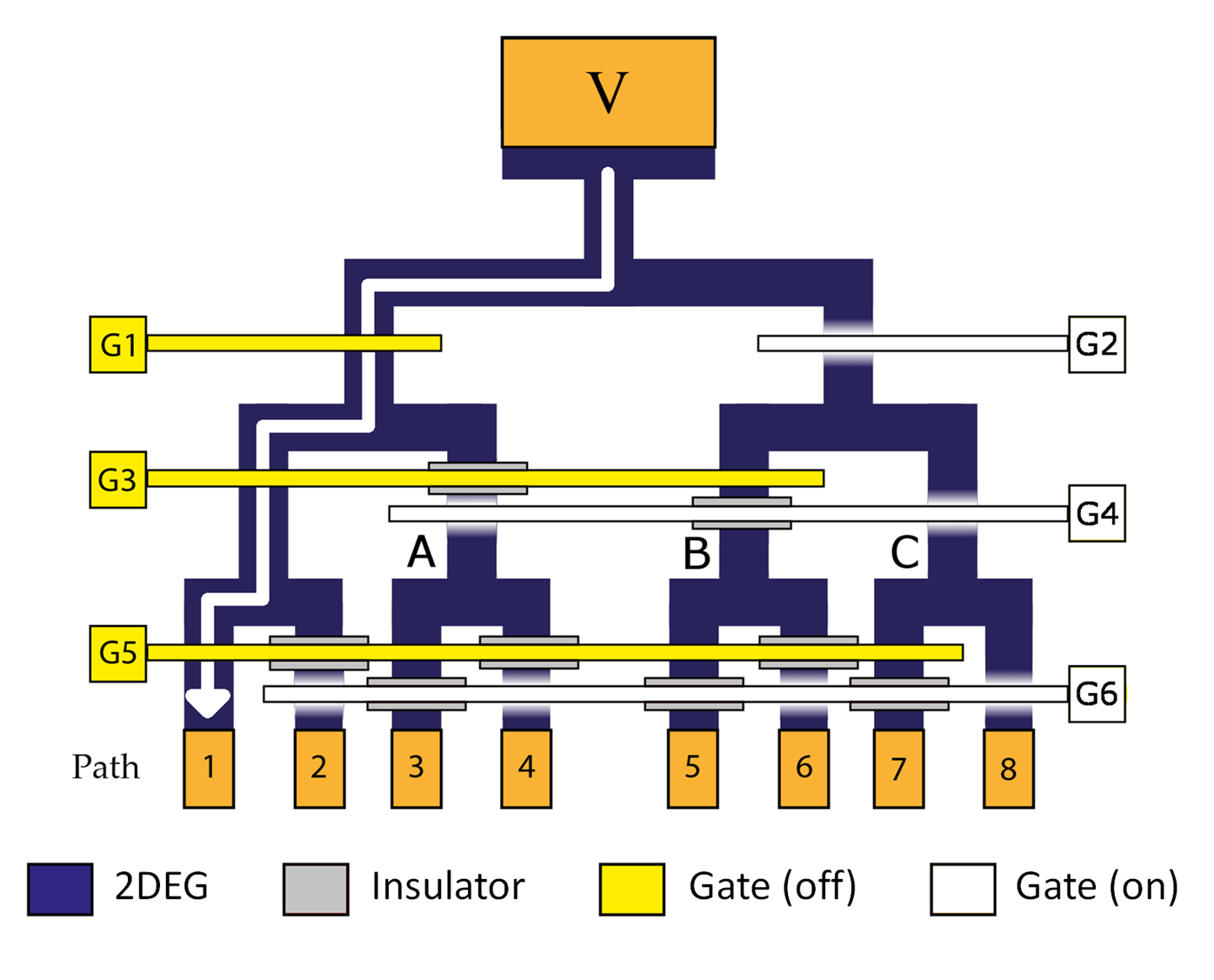}
\caption{\label{Fig1} (Color online) Schematic diagram of the multiplexer structure. The 2DEG (blue) branches from an input $V$ to eight separate outputs (labelled $1$ to $8$). Addressing gates (yellow and white) cross the 2DEG and are insulated in various places (grey). Addressing gates $G2$, $G4$ and $G6$ are turned `on' in order to direct an input voltage from $V$ to output path $1$, as illustrated by the arrow.}
\end{figure}

Figure~\ref{Fig1} shows a schematic diagram which illustrates the operation of the multiplexer in directing a voltage from input $V$ to one of eight output paths. A mesa is defined using standard etching techniques, and forms a tiered structure proceeding from input $V$. The path of the input voltage through the multiplexer is determined by the 6 `addressing gates' ($G1$ to $G6$), by negatively biasing the addressing gates to deplete carriers from the 2DEG below, thus preventing current flow through a particular arm of the multiplexer.

The addressing gates cover multiple arms of the multiplexer, for example, $G4$ covers the 2DEG at points $A$, $B$, and $C$. In order for the addressing scheme to be effective, path $B$ should remain open whilst paths $A$ and $C$ are depleted. Therefore, an insulator was deposited at $B$ between the addressing gate and the surface of the wafer. This alters the voltage required to deplete carriers in the 2DEG, and thus opens a voltage window where carriers are depleted at $A$ and $C$ but not at $B$. A 400 nm layer of photodefinable insulator, polyimide (HD4104), was used, which shifts the voltage required to deplete carriers from $-0.2$ V to $<$ $-5$ V.

In Fig.~\ref{Fig1}, gates $G2$, $G4$ and $G6$ are `on', i.e. a negative voltage is applied to deplete the 2DEG where there is no insulator. Gates $G1$, $G3$ and $G5$ are `off'. The input voltage at $V$ is therefore directed to path $1$. This voltage can be output on any of the other paths depending on which combination of addressing gates are on or off. The total number of output paths of the multiplexer increases approximately exponentially with each tier of the structure, and is given by $2^{(n-1)/2}$, where $n$ is the total number of contacts required (including the addressing gates and the input contact).

\begin{figure*}
\includegraphics[width=18cm,height=17.5cm,keepaspectratio]{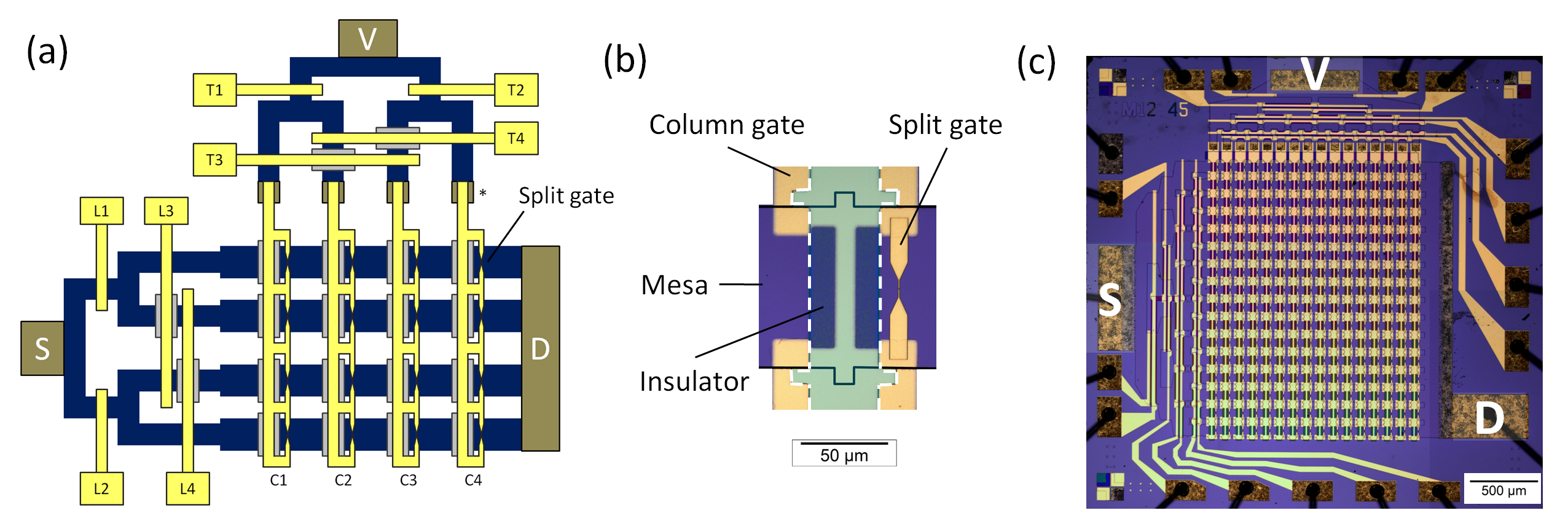}% Here is how to import EPS art
\caption{\label{Fig2} (Color online) (a) Simplified schematic diagram of the device layout, for an array of $4 \times 4$ split gates. Two multiplexers are used. The left-hand multiplexer (addressing gates $L1$ to $L4$) directs an ac excitation voltage from source $S$ to drain $D$ through one of the four mesa rows. The top multiplexer (addressing gates $T1$ to $T4$) directs an input voltage $V$ to columns $C1$ to $C4$. Individual split gates are measured by appropriately addressing both multiplexers to select a particular row and column. (b) Optical micrograph of one of the split gates in the array. For clarity, the edge of the mesa (insulator) is marked by the black (white-dotted) line, and the insulator beneath the column gate is highlighted (light blue, artificial colouring). (c) Optical micrograph of the array of 256 split gates ($16$ rows and $16$ columns). The source, drain, and input-voltage contacts are labelled $S$, $D$, and $V$, respectively. All gates are shown in yellow.}
\end{figure*}

Figure~\ref{Fig2}(a) shows a schematic diagram of the layout used to address an array of split gates, for illustrative purposes we show an array of $4 \times 4$ split gates. Two multiplexers are required; one to select the desired row (mesa), and one to select the column (gate). The ac excitation voltage is applied at source $S$ and is directed to a common drain contact $D$ through one of the four rows using the left-hand multiplexer (for which the addressing gates are labelled $L1$ to $L4$).

A series of ohmic contacts are positioned at the outputs of the top multiplexer. Columns $C1$ to $C4$ cover the ohmic contacts (for example at point $*$ for column $C4$). Thus, by selectively biasing addressing gates $T1$ to $T4$, the input voltage is directed from contact $V$ to a particular column. Both arms of the split gate are connected to the same column. Figure~\ref{Fig2}(b) shows an optical micrograph of one of the split gates in the array. The edges of the mesa are outlined by thick black lines for clarity. An insulating layer of polyimide prevents the 2DEG from being depleted beneath the column when a bias voltage is applied. The insulator is outlined by the white-dotted lines.
 
Adding an extra tier to either multiplexer doubles the number of split gates that can be contacted, at the expense of two additional addressing gates. We used the layout described to address an array of 256 split gates; currently the largest number of split gates that have been individually measured on a single chip.
An optical micrograph of the entire chip is shown in Fig.~\ref{Fig2}(c). The contacts for the source, drain and split-gate voltage ($V_{sg}$) are labelled $S$, $D$, and $V$, respectively. The chip was designed to fit into a standard 20-pin LCC package, to be compatible with the existing cryostat set-up. Nineteen contacts were required: 16 addressing gates (8 for each multiplexer), 2 contacts for the source and drain, and 1 contact for $V_{sg}$.

The conductance ($G$) through each split gate in the array was measured as a function of $V_{sg}$. Fifteen of the split gates failed to define a 1D channel, and it was found that in each case this was due to damage to one arm of the split gate incurred during fabrication. We therefore define a fabrication-limited yield ($Y_f$) of $Y_f=94\%$. 

It is also necessary to define a `quantum yield' ($Y_q$), since quantum phenomena in mesoscopic devices are affected by disorder. This disorder can arise from the presence of impurities in the wafer, fluctuations in the background potential due to ionized donors and surface states, and non-uniformities in the gate geometry. The specific definition of $Y_q$ will depend on the type of mesoscopic device, and its particular application.
In the case of split gates, the quantum phenomenon of interest is the quantisation of conductance in units of $2e^2/h$. Therefore, we define $Y_q$ as the number of devices for which the first and second conductance plateaux are clearly defined and occur at the correct conductance value, $2e^2/h$ and $4e^2/h$, respectively.
Due to the volume of data to be analysed, an algorithm was written to extract the value of the first and second plateaux, after correcting for series resistance. Practically, $Y_q$ was defined as the number of devices for which both conductance plateaux occurred within $\pm0.1 \times (2e^2/h)$ of the expected value.

The split-gate array was measured before and after illumination, referred to as the dark and light measurement, respectively. In the dark, $Y_q=54.8\%$, giving a total yield $Y_t=Y_f \times Y_q = 51.6 \%$.
In the light, $Y_q$ increased to $86.3\%$, giving $Y_t = 81.3 \%$.
In HEMT structures, illumination releases electrons trapped in DX centres in the donor layer~\cite{Lang1979}, giving rise to an increased carrier density and mobility in the 2DEG.
The 1D subband spacing increases with $n$, and therefore plateaux become better defined. Disorder effects are also reduced since the donor layer is fully ionized and thus gives rise to a more uniform background potential, and impurities are better screened due to the higher $n$.

The dependence of $Y_q$ on $n$ and $\mu$ indicate that a higher yield may be obtained by fabricating the split-gate array on a higher quality GaAs/AlGaAs HEMT. 
It is also likely that a higher yield can be achieved by laterally shifting the 1D channel away from impurities, thereby reducing the effect on disorder on 1D quantisation~\cite{Williamson1990}. This requires separate control over each arm of the split gate, which could be achieved with modifications of the multiplexer design.

\begin{figure}
\includegraphics[width=8.5cm,height=15cm,keepaspectratio]{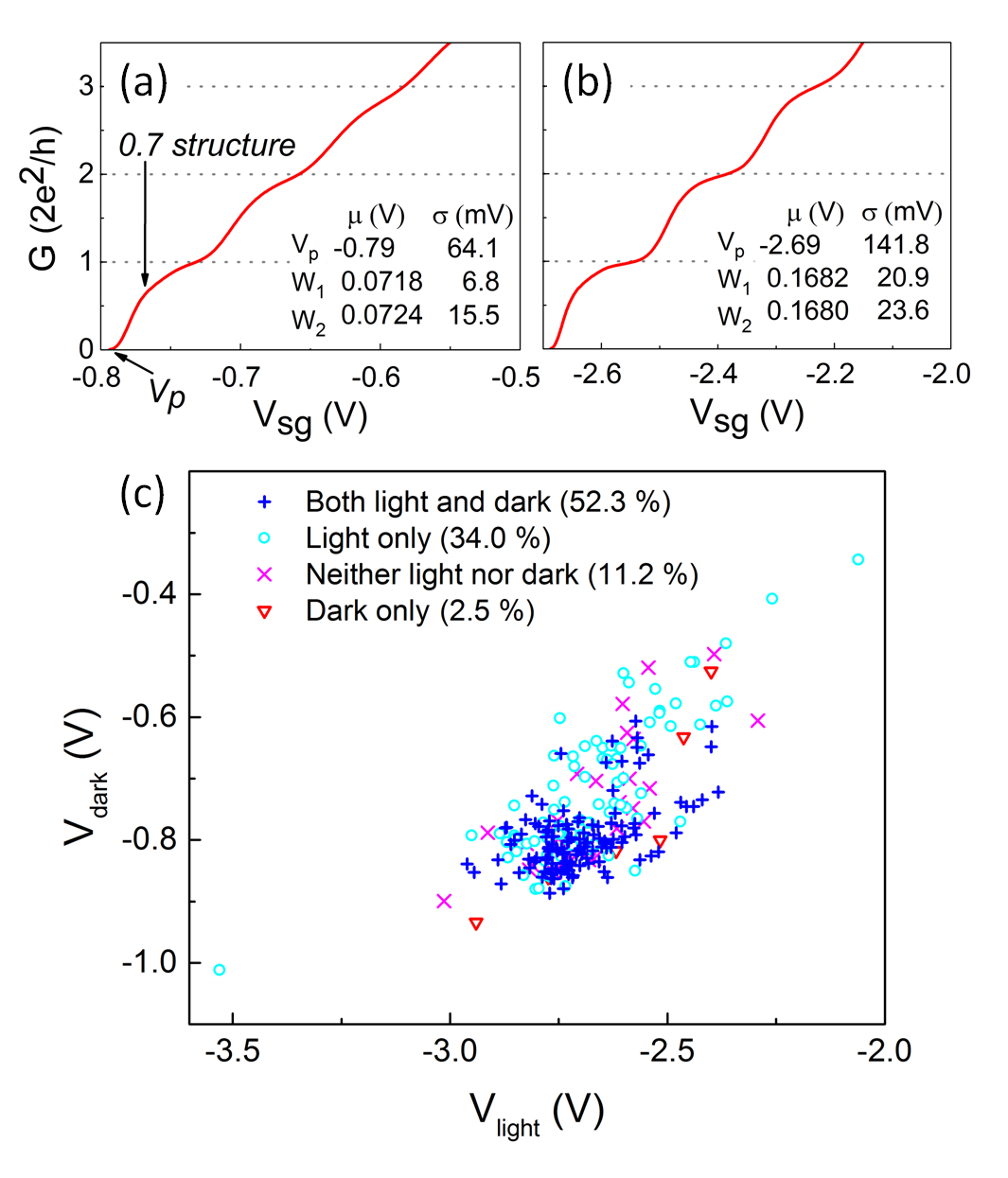}% Here is how to import EPS art
\caption{\label{Fig3} (Color online) (a) Average conductance $G$ against $V_{sg}$ for all split gates which met the yield criteria before illumination. The 0.7 structure and $V_p$ are indicated by the arrows. (b) Corresponding $G$ against $V_{sg}$ for split gates which met the yield criteria after illumination. For (a) and (b), the mean ($\mu$) and standard deviation ($\sigma$) for $V_p$, $W_1$ and $W_2$ are given in the inset, where $W_1$ and $W_2$ correspond the widths of the first and second plateaux, respectively. (c) Scatter plot of $V_{dark}$ against $V_{light}$ for all split gates, where $V_{dark}$ ($V_{light}$) is the pinch-off voltage before (after) illumination. The symbols indicate whether the devices met the quantum yield criteria in both the light and dark; neither the light nor dark; only in the light; or only in the dark.}
\end{figure}

Figure \ref{Fig3}(a) shows the average conductance $G$ against $V_{sg}$ for all split gates which met the yield criteria before illumination. The conductance data were shifted to the mean pinch-off voltage ($V_p$) before averaging, where $V_p$ is defined as the voltage at which $G=0$ [marked on Fig. \ref{Fig3}(a)]. The mean $V_{p} = -0.79$ V, and the mean widths of the first ($W_1$) and second ($W_2$) plateaux are $W_1=W_2=72$ mV (the plateau widths were approximated as the distance between the mid-points of the risers between conductance plateaux).
A weak `0.7 structure' also exists, indicated by the arrow. This structure is a feature of conductance through 1D devices, which appears close to $0.7 \times (2e^2/h)$~\cite{Thomas1996, Micolich2011}.
Figure \ref{Fig3}(b) shows the corresponding average conductance against $V_{sg}$ after illumination. The mean $V_p$ has shifted to $-2.69$ V, and the plateau widths have increased by a factor of $2.3$, reflecting the increase in $n$ and larger 1D subband spacing.

Figure \ref{Fig3}(c) shows a scatter plot of $V_p$ in the dark ($V_{dark}$) against $V_p$ in the light ($V_{light}$), for every split gate which defined a 1D channel. 
The data points are represented by one of four symbols to indicate which devices met the quantum yield criteria in both the light and dark ($52.3 \%$); in neither the light nor dark ($34.0 \%$); only in the light ($11.2 \%$); and only in the dark ($2.5 \%$).

There is a strong correlation between $V_{dark}$ and $V_{light}$.
The spread in the pinch-off voltages can be related to a variety of causes, including fluctuations in the background potential due to donor ions, and variations in the lithographic geometry. 
The standard deviation ($\sigma$) of $V_{dark}$ and $V_{light}$ is $97$ mV and $144$ mV, respectively, which corresponds to $12.7 \%$ and $5.4 \%$ of the mean~\cite{mean}.
The reduction in the percentage variation in the light may reflect greater uniformity in the ionized donor distribution following the release of electrons from DX centres. 
Studies can be imagined to determine how significant the different contributing factors are to the variation in $V_p$. For example, the role of ionized donors may be investigated by fabricating the sample on an undoped GaAs/AlGaAs heterostructure~\cite{Kane1993}, where the absence of dopants reduces the background disorder potential~\cite{See2012}. It would be necessary to modify the multiplexer design to include a gate to induce the 2DEG~\cite{Harrell1999}.

Measuring 256 split gates on a single chip provides a data set sufficient for statistical analysis. The only investigation of characteristics of a large number of split gates was conducted by Yang~\emph{et al.}~\cite{Yang2009}, who measured a total of 540 devices. However, only $6$ devices were fabricated on each GaAs/AlGaAs chip, therefore 90 cooldowns were required to obtain the data. By fabricating 256 split gates on a single chip, we only require two cooldowns to produce a similar volume of data.

This demonstrates a major advantage of the multiplexing scheme, in that ordinarily, the number of devices that can be measured on a single chip is limited by the number of wires in a cryostat. The multiplexer effectively increases the number of electrical contacts available, such that existing experimental set ups do not need to be modified.
Fewer cooldowns are required to measure a large number of devices, therefore more data can be gathered in a shorter amount of time, at a fraction of the cost. Additionally, measuring a large number of devices increases the likelihood of encountering rare situations where impurities within the active region of the device lead to unusual quantum effects, which can be investigated.

In summary: A multiplexing scheme has been developed which dramatically increases the number of mesoscopic devices that can be measured on a single GaAs/AlGaAs chip. The multiplexer was used to quantify the yield of an array of 256 split gates, where the definition of yield includes the effects of disorder on quantisation of conductance. The maximum total yield for the split-gate array was $81.3 \%$, which can be improved by increasing the mobility and carrier density of the 2DEG. The data set obtained is sufficient to investigate statistics of complex quantum phenomena, which will be the subject of a future study. Additionally, the multiplexer is a scalable architecture, which can be extended to other mesoscopic devices with some modifications. Investigating the yield and statistical variations of quantum phenomenon in mesoscopic devices, as well as demonstrating a scalable device architecture is necessary in testing the suitability of these devices for computational applications. During the preparation of this manuscript we became aware of the work of Ward~\emph{et al.}~\cite{Ward2013}, which presents an alternative multiplexing scheme used to contact four double quantum dot structures on a Si/SiGe heterostructure.

%\begin{acknowledgments}
This work was supported by the Engineering and Physical Sciences Research Council Grant No. EP/I014268/1.\\ The authors would like to thank C. J. B. Ford, I. Farrer and F. Sfigakis for invaluable discussions, and R. D. Hall for electron-beam exposure.
%\end{acknowledgments}

%\bibliography{aipsamp}% Produces the bibliography via BibTeX.

\end{document}